\documentclass[aoas,MSNbibl,nameyear,dvips]{arximspdf}
\usepackage{graphicx}
\doi{10.1214/12-AOAS577} %kopijuoti is PTS
\volume{7}
\issue{1}
\pubyear{2013}
\firstpage{249}
\lastpage{268}



\begin{document}
\begin{frontmatter}

\title{Robust partial
likelihood approach for detecting imprinting and maternal effects using
case-control~families\thanksref{T1}}
\runtitle{LIME for case-control family}

\thankstext{T1}{Supported in part by NSF Grant DMS-12-08968.}

\begin{aug}
\author[A]{\fnms{Jingyuan} \snm{Yang}}
\and
\author[A]{\fnms{Shili} \snm{Lin}\corref{}\ead[label=e2]{shili@stat.osu.edu}}
\runauthor{J. Yang and S. Lin}
\affiliation{Ohio State University}
\address[A]{Department of Statistics\\
Ohio State University\\
1958 Neil Ave\\
Columbus, Ohio 43210\\
USA\\
\printead{e2}}
\end{aug}

% HISTORY:
\received{\smonth{3} \syear{2011}}
\revised{\smonth{2} \syear{2012}}

% ABSTRACT
%
\begin{abstract}
Genomic imprinting and maternal effects are two epigenetic factors that
have been increasingly explored for their roles in the etiology of
complex diseases. This is part of a concerted effort to find the
``missing heritability.'' Accordingly, statistical methods have been
proposed to detect imprinting and maternal effects simultaneously based
on either a case-parent triads design or a case-mother/control-mother
pairs design. However, existing methods are full-likelihood based and
have to make strong assumptions concerning mating type probabilities
(nuisance parameters) to avoid overparametrization. In this paper we
propose to augment the two popular study designs by combining them and
including control-parent triads, so that our sample may contain a
mixture of case-parent/control-parent triads and
case-mother/control-mother pairs. By matching the case families with
control families of the same structure and stratifying according to the
familial genotypes, we are able to derive a partial likelihood that is
free of the nuisance parameters. This renders unnecessary any
unrealistic assumptions and leads to a robust procedure without
sacrificing power. Our simulation study demonstrates that our partial
likelihood method has correct type I error rate, little bias and
reasonable power under a variety of settings.
%Although methods that make assumptions
%about the nuisance parameters may, albeit not always,
% lead to higher power, such gains are in the expense of greatly
%inflated type I error rate and bias should the assumptions be violated.
\end{abstract}

% KEYWORDS
%

\end{frontmatter}

%s1 #&#
\section{Introduction}
%What are IP and ME and why they are important
Genomic imprinting and maternal effects are two epigenetic factors that
have been increasingly
explored for their roles in the etiology of complex diseases.
This is part of a concerted effort to find the ``missing heritability''
[\citet{Manolio2009}].
Genomic imprinting (maternal or paternal) is an epigenetic process
involving methylation and histone modifications in order to silence the
expression of a gene inherited from a particular
parent (mother or father) without altering the genetic sequence.
This process leads to unequal expression of a heterozygous genotype
depending on
whether the imprinted variant is inherited from the mother or the father.
Maternal effect, on the other hand,
refers to a situation where the phenotype of an individual is
influenced by
the genotype of the mother.
Maternal effects usually occur due to the additional mRNAs or proteins
passed from mother to fetus during pregnancy, which may result in an
individual showing
the phenotype due to the genotype of the mother regardless of one's own
genotype.

%Diseases associated with IP and MA
The first imprinted gene in humans was found almost 20 years
ago
[Giannoukakis et~al. (\citeyear{Giannoukakis1993})].
Since then, a number of genetic disorders have been found to be
associated with
imprinting defects. The most well known include
Beckwith--Wiedemann syndrome, Silver--Russell syndrome, Angelman
syndrome and
Prader--Willi syndrome [\citet{Falls}].
Although it has been estimated that about 1\% of all mammalian genes
are imprinted
[\citet{Morison}], only a limited number have been identified
thus far.
With the availability of the massively parallel sequencing technology,
scientists
are now able to carry out direct studies of imprinting genomewide in
mouse efficiently
[\citet{Gregg2010}, \citet{Wang2008}]. Nevertheless, the controlled
mating setup that was successful in
mouse studies is not feasible in humans. Therefore,
there is still a pressing need for the development of robust and powerful
statistical methods for detecting
imprinting effects based on single nucleotide polymorphism data.

A variety of diseases, especially those that are related to pregnancy
outcomes, such as childhood diseases and birth defects, have been
hypothesized to
be influenced by maternal effects. Well-known examples of
diseases in which maternal effects play a role include childhood cancer
and spina bifida [Haig (\citeyear{Haig1993}, \citeyear{Haig2004}), \citet{Jensen2006}].
Maternal effects are also suspected to be involved in other
diseases such as schizophrenia [\citet{Palmer2006}] and high blood
pressure [\citet{yangandlin}].
Although imprinting and maternal effects arise from two different
biological processes,
their effects as expressed in the phenotypes can mask one another
[\citet{mimic}].
Thus, it is important that these two confounding effects be studied
jointly to avoid
false positives/negatives.

%Two popular designs
Two popular designs for studying genomic imprinting and/or maternal
effects are case-parent triads and case-mother pairs, the latter may be
supplemented by control-mother pairs [\citet{Weinberg1998},
\citet{Weinberg2005}, \citet{casecontrol},
\citet{Ainsworth2011}]. The use of triads is attractive, as both
imprinting and maternal effects can be studied jointly to handle the
issue of confounding. However, it is well known that fathers are
usually much harder to recruit than mothers for a genetic study, and
thus a study design with case-mother/control-mother pairs may be easier
to meeting its target sample size. Nevertheless, the pair design will
lead to reduction in the number of familial genotype categories, and
thus only maternal effect\vadjust{\goodbreak} is usually studied even with the assumption
of mating symmetry [\citet{casecontrol}]. The hybrid design of
\citet{Vermeulen2009} proposed to use control-mother pairs to
enrich the sample of case-parent triads, leading to a case-parent
triad/control-mother design. Such a design increases the number of
family genotype categories so that mating frequencies can be estimated
without the mating symmetry assumption.
%Nevertheless, \citet{Vermeulen2009} only considered
%maternal effects assuming there is no imprinting, although imprinting
%effect may be incorporated following \citet{Weinberg1999}.}

%Existing methods
Both nonparametric and parametric statistical methods
have been proposed in the literature for analyzing triads and pairs data.
Nonparametric methods are mainly for detecting the imprinting
effect under the assumption of no maternal effect [\citet{Weinberg1999},
\citet{jiyuanCI}]. Such methods are
very attractive, as they are simple,
elegant and powerful, but they may suffer from severely inflated
type I error rate or power loss if the
assumption is violated.
On the other hand, existing parametric methods can usually study both imprinting
and maternal effects (with triad data), but they usually need to make rather
strong assumptions to avoid overparametrization.
The typical assumptions made are regarding
mating type probabilities, the most extreme of which is random mating,
leading to the
Hardy--Weinberg equilibrium (HWE). More mild assumptions include
parental allelic
changeability and mating symmetry.
It is well accepted that the assumption of HWE is unlikely to hold, however,
even the validity of the least stringent one, mating symmetry, may
still be in doubt in
reality.
%For example, men with brown eyes might prefer to marry women with blue
%eyes, but the reverse may not
%be true. As such, the mating type of brown-eye men with blue-eye women
%may not be the same as that of brown-eye women with blue-eye men,
%violating
%the mating symmetry assumption.
Mating selections are usually guided by cultural values and social
rules in general, and, further, mating symmetry is easily violated if
there is any gender-specific assortative mating in the population.
%Despite the doubt about mating type probability assumptions, they are
%sometimes made to
%increase the statistical power \cite{casecontrol} in the expense of
%inflated type I error
%rates had such assumptions not hold. However, the most important
%reason for making such unrealistic assumptions is
%technical: some models would otherwise be over-parametrized if such
%assumptions are not
%made.

% rare vs. common diseases
Many complex diseases, such as cancers and hypertension, are rather
common. However,
rare diseases are sometimes assumed in existing methods so that
the probabilities of child-mother pair genotype combinations
can be treated as approximately the
same in the control and in the general populations [\citet{casecontrol}].
Although the rationale seems rather intuitive, analytical as well as simulation
results indicate that the rare disease assumption is in fact only a
necessary, but not a sufficient
condition, for the frequencies to be roughly equal. %The allele
%frequencies play a
%crucial role, in addition to the rare disease assumption,
It is the interplay of allele frequency and the underlying genetic
model, not
the rare disease assumption alone, that determines whether
the pair frequencies are roughly equal.

%What we plan to do in this paper.
In this paper, we propose a partial \textit{L}ikelihood approach for
detecting \textit{I}mprinting and
\textit{M}aternal \textit{E}ffects (LIME) simultaneously using a family-based
case-control design.
Specifically, our sampled data is a
mixture of case-parent/control-parent triads and
case-mother/control-mother pairs.
By including control families in our design, we can match case families
with control families of the same structure (i.e., triads vs. triads
and pairs
vs. pairs) and factor out common terms involving mating type
probabilities, the
nuisance parameters.
This design makes it possible to formulate a novel partial likelihood approach,
wherein the likelihood component of interest is free of the nuisance parameters.\vadjust{\goodbreak}
This circumvents the problem of overparametrization and unrealistic
assumptions that
plague existing methods.
Further, LIME can make use of all available data information (complete
triads and those
with missing fathers). It does not rely on any assumption about mating type
probabilities or rarity of the disease, which makes it robust to
departure from
the usual assumptions without sacrificing much power.\looseness=-1

%In this paper, we propose a partial likelihood approach for detecting
%imprinting and maternal effects (LIME) simultaneously using a
%family-based case-control design. Specifically, we consider nuclear
%families with a single child who is either affected or unaffected, and
%the father is either available or unavailable. Thus, our sampled data
%is a
% a mixture of case-parent/control-parent triads and
%case-mother/control-mother pairs. We assume multiplicative relative
%risk due to the variant allele, imprinting and maternal effects, and
%model the likelihood of the counts of case families and control
%families with specific genotype combinations. The recruited control
%families can be compared with case families for estimating the disease
%odds. More importantly, through matching the familial genotype
%combinations of case and control families, nuisance parameters of
%mating type frequencies in the population can be eliminated, and only
%the partial likelihood involving the parameters of genotype relative
%risks needs to be considered.
%This circumvents the problem of over-parametrization and unrealistic
%assumptions that plague existing methods.
%In other words, LIME can make use of all available data information
%(triads or those
%with missing fathers). It does not rely on any assumption about mating
%type
%probabilities or the disease being rare, making it much more robust to
%departure from
%the usual assumption without sacrifycing much power.

%s2 #&#
\section{Method}

The genotype scores (the number of the variant alleles carried by an
individual) of the mother, father and child in a triad are denoted by
$M$, $F$ and~$C$, respectively, which takes values in $\{0, 1, 2\}$.
In case-mother or control-mother pairs, the paternal genotype score $F$
is missing. The disease status $D=1$ indicates that the child is
affected and 0 otherwise. We use the following multiplicative relative
risk model of disease penetrance: % proposed by \cite{weinberg1998}:
%
%e1 #&#
%
\begin{eqnarray}
\label{mm}
&&
P(D=1|M,F,C)\nonumber\\[-8pt]\\[-8pt]
&&\qquad=\delta R_1^{I(C=1)}
R_2^{I(C=2)} R_{{im}}^{I(C=1\ \mathrm{and}
\ \mathrm{origin}=M)} S_1^{I(M=1)}
S_2^{I(M=2)},\nonumber
\end{eqnarray}
where $\delta$ is the phenocopy rate of the disease; $R_1$ and $R_2$
are the variant allele effect of 1 and 2 copies carried by the child,
respectively; $R_{{im}}$ is the effect when the single copy of the
variant allele carried by the child is inherited from the mother; $S_1$
and $S_2$ are the maternal effect when the mother carries 1 and 2
copies of the variant allele, respectively; and $I(\cdot)$ is the usual
indicator function that is equal to 1 or 0 depending on whether the
condition within the parentheses is met or not. Likelihood ratio tests
can be conducted to test: (1) for association, $H_0\dvtx
R_1=R_2=R_{{im}}=S_1=S_2=1$ vs. $H_a$: at least one of these parameters
is not~1; (2)~for (maternal/paternal) imprinting, $H_0\dvtx R_{{im}}=1$
vs. $H_a\dvtx R_{{im}}(</>)\neq1$; (3) for maternal effect, $H_0\dvtx
S_1=S_2=1$ vs. $H_a\dvtx S_1$ or $S_2$ is not 1. Note that this model
was used by \citet{Weinberg1998}, which was adopted by
\citet{casecontrol} but without the term for the imprinting effect
for their pair sampling design. This model is also equivalent to one of
the models used recently by \citet{Ainsworth2011}.

Denote the numbers of case-parent triads, control-parent triads,
case-mother pairs and control-mother pairs in the sample by $N_t^1$,
$N_t^0$, $N_p^1$ and $N_p^0$, respectively. Two special cases of this
general setup are as follows: (1) $N_p^0=N_p^1=0$, if there are no
missing fathers and thus all the families are triads; and (2)
$N_t^0=N_t^1=0$, if all fathers are missing, and thus all the families
are child-mother pairs. Therefore, our method represents a unified
approach as it is applicable to triad only data as in
\citet{Weinberg1998}, or pairs only data as in
\citet{casecontrol}, or a mixture of the two.
%Denote the counts of case-parent and control-parent triads with $M=m$,
%$F=f$ and $C=c$ by $n^1_{mfc}$ and $n^0_{mfc}$, respectively.
%Similarly, the count of case-mother and control-mother pairs with
%$M=m$ and $C=c$ are denoted by $n^1_{mc}$ and $n^0_{mc}$.

%our method
%
%t1 #&#
%
\begin{table}
\caption{Joint probabilities of disease status and triad genotypes\protect\tabnoteref{ta}}
\label{tabl1}
\begin{tabular*}{\tablewidth}{@{\extracolsep{\fill}}lcccll@{}}
\hline
\textbf{Type}
& $\bolds{M}$ & $\bolds{F}$
& $\bolds{C}$ & \multicolumn{1}{c}{$\bolds{P(D=1,M,F,C)}$}
& \multicolumn{1}{c@{}}{$\bolds{P(D=0,M,F,C)}$}\\
\hline
\hphantom{0}1&0&0&0&
$\mu_{00}\cdot1\cdot\delta$&
$\mu_{00}\cdot1\cdot[1-\delta]$\\[2pt]
\hphantom{0}2&0&1&0&
$\mu_{01}\cdot\frac{1}{2}\cdot\delta$&
$\mu_{01}\cdot\frac{1}{2}\cdot[1-\delta]$\\[2pt]
\hphantom{0}3&0&1&1&
$\mu_{01}\cdot\frac{1}{2}\cdot\delta R_1$&
$\mu_{01}\cdot\frac{1}{2}\cdot[1-\delta R_1]$\\[2pt]
\hphantom{0}4&0&2&1&
$\mu_{02}\cdot1\cdot\delta R_1$&
$\mu_{02}\cdot1\cdot[1-\delta R_1]$\\[2pt]
\hphantom{0}5&1&0&0&
$\mu_{10}\cdot\frac{1}{2}\cdot\delta S_1$&
$\mu_{10}\cdot\frac{1}{2}\cdot[1-\delta S_1]$\\[2pt]
\hphantom{0}6&1&0&1&
$\mu_{10}\cdot\frac{1}{2}\cdot\delta S_1R_1R_{{im}}$&
$\mu_{10}\cdot\frac{1}{2}\cdot[1-\delta S_1R_1R_{{im}}]$\\[2pt]
\hphantom{0}7&1&1&0&
$\mu_{11}\cdot\frac{1}{4}\cdot\delta S_1$&
$\mu_{11}\cdot\frac{1}{4}\cdot[1-\delta S_1]$\\[2pt]
\hphantom{0}8&1&1&1&
$\mu_{11}\cdot\frac{1}{4}\cdot\delta S_1 R_1 (1+R_{{im}})$&
$\mu_{11}\cdot\frac{1}{4}\cdot[2-\delta S_1 R_1 (1+R_{{im}})]$\\[2pt]
\hphantom{0}9&1&1&2&
$\mu_{11}\cdot\frac{1}{4}\cdot\delta S_1R_2$&
$\mu_{11}\cdot\frac{1}{4}\cdot[1-\delta S_1R_2]$\\[2pt]
10&1&2&1&
$\mu_{12}\cdot\frac{1}{2}\cdot\delta S_1R_1$&
$\mu_{12}\cdot\frac{1}{2}\cdot[1-\delta S_1R_1]$\\[2pt]
11&1&2&2&
$\mu_{12}\cdot\frac{1}{2}\cdot\delta S_1R_2$&
$\mu_{12}\cdot\frac{1}{2}\cdot[1-\delta S_1R_2]$\\[2pt]
12&2&0&1&
$\mu_{20}\cdot1\cdot\delta S_2R_1R_{{im}}$&
$\mu_{20}\cdot1\cdot[1-\delta S_2R_1R_{{im}}]$\\[2pt]
13&2&1&1&
$\mu_{21}\cdot\frac{1}{2}\cdot\delta S_2R_1R_{{im}}$&
$\mu_{21}\cdot\frac{1}{2}\cdot[1-\delta S_2R_1R_{{im}}]$\\[2pt]
14&2&1&2&
$\mu_{21}\cdot\frac{1}{2}\cdot\delta S_2R_2$&
$\mu_{21}\cdot\frac{1}{2}\cdot[1-\delta S_2R_2]$\\[2pt]
15&2&2&2&
$\mu_{22}\cdot1\cdot\delta S_2R_2$&
$\mu_{22}\cdot1\cdot[1-\delta S_2R_2]$\\\hline
&&&&$\bolds{P(D=1, M,C)}$ & $\bolds{P(D=0, M,C)}$ \\
\hline
1, 2&0&$-$&0&
$(\mu_{00}+\frac{1}{2}\mu_{01})\cdot\delta$&
$(\mu_{00}+\frac{1}{2}\mu_{01})\cdot[1-\delta]$\\[2pt]
3, 4&0&$-$&1&
$(\frac{1}{2}\mu_{01}+\mu_{02})\cdot\delta R_1$&
$(\frac{1}{2}\mu_{01}+\mu_{02})\cdot[1-\delta R_1]$\\[2pt]
5, 7&1&$-$&0&
$(\frac{1}{2}\mu_{10}+\frac{1}{4}\mu_{11})\cdot\delta S_1$&
$(\frac{1}{2}\mu_{10}+\frac{1}{4}\mu_{11})\cdot[1-\delta S_1]$\\[2pt]
6, 8, 10&1&$-$&1&
$\frac{1}{2}\mu_{10}\cdot\delta S_1R_1R_{{im}}$&
$\frac{1}{2}\mu_{10}\cdot[1-\delta S_1R_1R_{{im}}]$\\[2pt]
&&&&
$\quad{}+\frac{1}{4}\mu_{11}\cdot\delta S_1R_1(1+R_{{im}})$&
$\quad{}+\frac{1}{4}\mu_{11}\cdot[2-\delta S_1R_1(1+R_{{im}})]$\\[2pt]
&&&&
$\quad{}+\frac{1}{2}\mu_{12}\cdot\delta S_1R_1$&
$\quad{}+\frac{1}{2}\mu_{12}\cdot[1-\delta S_1R_1]$\\[2pt]
9, 11&1&$-$&2&
$(\frac{1}{4}\mu_{11}+\frac{1}{2}\mu_{12})\cdot\delta S_1R_2$&
$(\frac{1}{4}\mu_{11}+\frac{1}{2}\mu_{12})\cdot[1-\delta S_1R_2]$\\[2pt]
12, 13&2&$-$&1&
$(\mu_{20}+\frac{1}{2}\mu_{21})\cdot\delta S_2R_1R_{{im}}$&
$(\mu_{20}+\frac{1}{2}\mu_{21})\cdot[1-\delta S_2R_1R_{{im}}]$\\[2pt]
14, 15&2&$-$&2&
$(\frac{1}{2}\mu_{21}+\mu_{22})\cdot\delta S_2R_2$&
$(\frac{1}{2}\mu_{21}+\mu_{22})\cdot[1-\delta S_2R_2]$\\\hline
\end{tabular*}
\tabnotetext[a]{ta}{$M$, $F$ and $C$ are the number of variant allele(s)
carried by the mother, the father and the child in a family, which are
equal to 0, 1 or 2; $F=-\mbox{ indicates}$ paternal genotype is missing in
case-mother and control-mother pairs; $\mu_{mf}$ denotes the
probability of parental pairs in which the mothers carry $m$ copies and
the fathers carry $f$ copies of the variant allele, that is, mating
type probability of $(M, F)=(m,f)$; $\delta$ is the phenocopy rate of
the disease in the population; $R_1$ and $R_2$ are relative risks due
to 1 and 2 two copies of the variant allele carried by the offspring,
respectively; $R_{{im}}$ is the relative risk due to the single copy of
the variant allele being inherited from the mother; $S_1$ and $S_2$ are
the maternal effect of 1 and 2 copies of the variant allele carried by
the mother, respectively.}
\end{table}

There are 15 possible combinations of $(M,F,C)$ for triads; their
enumeration and labeling
(type) are listed in the top segment of Table~\ref{tabl1}. The joint probability
of the
disease status of the child $(D)$ and the triad genotype combination
$(M,F,C)$ can be
decomposed as
\[
P(D,M,F,C)=P(M,F)P(C|M,F)P(D|M,F,C),
\]
where $P(M,F)$ is the population probability of a particular parental
pair (also called mating type), which is denoted by $\mu_{mf}$ for
$M=m$ and $F=f$. Note that we do not make any assumption about the
mating type probabilities such as HWE or even mating symmetry, and thus
$\mu_{mf}$ is not necessarily equal to $\mu_{fm}$. On the other hand,
$P(C|M,F)$ is the transmission probability, which follows the Mendelian
law of segregation. For the penetrance probability, $P(D|M,F,C)$, we
use the multiplicative relative risk model as given in equation (\ref{mm}).
For all types other than type 8 (Table~\ref{tabl1}), if a child has one
copy of the variant allele, the parental origin can be unambiguously
ascertained, and hence the relevant factors can be easily extracted
from (\ref{mm}) and multiplied to the joint probability. For type 8, in
which $(M,F,C)=(1,1,1)$, the variant allele carried by the child can be
inherited either from the mother or the father with equal probabilities
and, as such, both possibilities need to be considered, leading to the
summation of two probabilities weighted equally. For all 15 types, the
specific joint probabilities for the case-parent and control-parent
triads are given in the last two columns of the top segment of
Table~\ref{tabl1}. We can see from the table that the parameters
concerning the mating type probabilities $(\mu_{mf})$ are factored out
nicely from the parameters of the disease model, both for the case and
the control families.

For pairs, because the father's genotype is missing, the 15
$(M,F,C)$ combinations for triads collapse into 7 $(M,C)$ types, as
indicated in the bottom segment of Table~\ref{tabl1}. In other words, each
$(M,C)$ type is the combination of
potential triad types had the paternal genotype been observed for those
child-mother pairs.
%Since these combinations are unique for each $(M,C)$ type and
%therefore are treated as
%labels, as given under ``type'' in the bottom segment of Table
For example, for $(M,C)=(0,0)$, the
father's genotype can be either 0 or 1 and, therefore, the first type
for $(M,C)$
is the combinations of types 1 and 2 for triads. Accordingly, the joint
distribution of disease status and genotype is the sum of the
probabilities of the
collapsing triads types.
That is,
\[
P(D,M,C)=\sum_{F=0,1,2} P(D,M,F,C)=\sum
_{F=0,1,2} P(M,F,C) P(D|M,F,C),
\]
where $P(M,F,C)$ may be 0 if $F$ is not compatible with the observed
genotypes. For example, for $(M,C)=(0,0)$, $F=2$ is incompatible and,
therefore, the summation is only over 0 and 1.
For most $(M,C)$ combinations, the penetrance $P(D|M,F,C)$ can be
factorized out
of the summation over $F=0,1,2$, since the penetrance is the same for
the potential types
of triad regardless of the paternal genotype under our model.
The only exception is $(M,C)=(1,1)$, in which case the penetrance term is
different for the three potential types of triad unless the imprinting
effect is absent (i.e.,
$R_{{im}}=1$).
The specific joint probabilities for the case-mother and control-mother
pairs are given in the last two columns of the bottom segment of Table
\ref{tabl1}.
We can see from the table that, other than when $(M,C)=(1,1)$, the
parameters concerning the mating type probabilities $(\mu_{mf})$
are factored out nicely from the parameters of the disease model, both
for the case
and the control families.

Denote the counts of case-parent and control-parent triads with $M=m$,
$F=f$ and $C=c$ by $n^1_{mfc}$ and $n^0_{mfc}$, respectively.
Similarly, the counts of case-mother and control-mother pairs with
$M=m$ and $C=c$ are denoted by $n^1_{mc}$ and $n^0_{mc}$.
With the fixed totals $N_t^1$, $N_t^0$, $N_p^1$ and $N_p^0$, the
distributions of the
15 observed triad
counts for the cases $(\{n_{mfc}^1\})$ and the controls $(\{n_{mfc}^0\})$,
and those of the 7 observed pair counts for the cases $(\{n_{mc}^1\})$
and the controls
$(\{n_{mc}^1\})$, are as follows:
\begin{eqnarray*}
n^1_{mfc}&\sim& \operatorname{Multinomial}\bigl(N_t^1,
P(M=m,F=f,C=c|D=1)\bigr),
\\
n^0_{mfc}&\sim& \operatorname{Multinomial}\bigl(N_t^0,
P(M=m,F=f,C=c|D=0)\bigr),
\\
n^1_{mc}&\sim& \operatorname{Multinomial}\bigl(N_p^1,
P(M=m,C=c|D=1)\bigr),
\\
n^0_{mc}&\sim& \operatorname{Multinomial}\bigl(N_p^0,
P(M=m,C=c|D=0)\bigr).
\end{eqnarray*}
The cell probabilities $P(M,F,C|D)=P(D,M,F,C)/P(D)$ and
$P(M,C|D)=P(D,M,C)/P(D)$, where the probabilities in the numerators are
as given in
Table~\ref{tabl1} and we assume the disease prevalence, $P(D=1)$, in
the source
population is
known.
Such information for
most known diseases can be retrieved from the Incidence and Prevalence
Database (IPD)
(\href{http://www.tdrdata.com/IPD/ipd\_init.aspx}{http://www.tdrdata.com/}
\href{http://www.tdrdata.com/IPD/ipd\_init.aspx}{IPD/ipd\_init.aspx})
or other sources.

The likelihood function of our observed data is as follows:
\begin{eqnarray*}
&&
P\bigl(n_{mfc}^1, n_{mfc}^0,
n_{mc}^1, n_{mc}^0|
\mu_{mf}, \delta, R_1, R_2, R_{{im}},
S_1, S_2\bigr)
\\
&&\qquad\propto\prod
_{(m,f,c)}P(M=m,F=f,C=c|D=1)^{n_{mfc}^1}\\
&&\qquad\quad\hspace*{26.6pt}{}\times P(M=m,F=f,C=c|D=0)^{n_{mfc}^0}
\\
&&\qquad\quad{}\times \prod_{(m,c)\neq
(1,1)}P(M=m,C=c|D=1)^{n_{mc}^1}P(M=m,C=c|D=0)^{n_{mc}^0}
\\
%&=\prod_{(m,f,c)}\left[\frac{P(D=1|M=m,F=f,C=c)}{P(D=1)}
%& \prod_{(m,c)\neq(1,1)}\left[\frac{P(D=1|M=m,C=c)}{P(D=1)}
%M=m,C=c)}{P(D=0)}\right]^{n_{mc}^0}\\
%& \left\{\left[P(M=m,M=f,C=c)\right]^{n_{mfc}^1+n_{mfc}^0}
&&\qquad\propto\biggl\{\prod
_{(m,f,c)}(p_{mfc})^{n_{mfc}^1}(1-p_{mfc})^{n_{mfc}-n_{mfc}^1}\\
&&\hspace*{3pt}\qquad\quad{}\times\prod_{(m,c)\neq
(1,1)}(p_{mc})^{n_{mc}^1}(1-p_{mc})^{n_{mc}-n_{mc}^1}
\biggr\}
\\
&&\qquad\quad{}\times\biggl\{\prod_{(m,f,c)}\bigl[s_{mfc}P(M=m,F=f,C=c)
\bigr]^{n_{mfc}}\\
&&\hspace*{32pt}\qquad\quad{}\times\prod_{(m,c)\neq
(1,1)}\bigl[s_{mc}P(M=m,C=c)
\bigr]^{n_{mc}} \biggr\},
\end{eqnarray*}
where $n_{mfc}=n_{mfc}^0+n_{mfc}^1$ and $n_{mc}=n_{mc}^0+n_{mc}^1$.
Further,
\begin{eqnarray*}
s_{mfc}&=&N_t^1P(D=1|m,f,c)/P(D=1)+N_t^0P(D=0|m,f,c)/P(D=0),
\\
s_{mc}&=&N_p^1P(D=1|m,c)/P(D=1)+N_p^0P(D=0|m,c)/P(D=0),
\\
p_{mfc}&=&\frac{N_t^1P(D=1|m,f,c)/P(D=1)}{s_{mfc}},
\\
p_{mc}&=&\frac{N_p^1P(D=1|m,c)/P(D=1)}{s_{mc}}.
\end{eqnarray*}
Note that the above expressions are independent of the nuisance parameters
(mating type probabilities
$\mu_{mf}$), as we will see more clearly, from the derivation of the
formula below,
the nuisance parameters in the numerator
and denominator cancel out.

In our likelihood formulation, note that the pair type $(1,1)$ is not included
since the nuisance parameters and the risk parameters cannot be nicely separated
as discussed above. The potential effects of excluding this data type
will be addressed in the discussion section. With this exclusion, one
can see
from the above factorization that only the second component of the likelihood
(factors within the second set of
curly brackets in the last formula)
contains the nuisance parameters
$\mu_{mf}$'s.
Therefore, the first component is our partial likelihood,
which
can be maximized instead of the full likelihood to avoid estimating the nuisance
parameters [\citet{Cox1975}].
In fact, the partial likelihood component can be regarded as the
likelihood of
the reorganized data conditional on each possible triad $(M,F,C)$ or
pair $(M,C)$ type.
Within each type, counts of the cases and controls follow a ``renormalized''
binomial distribution with the appropriate probabilities. For example,
for a triad type $(m,f,c)$, the probabilities of observing a case are
\begin{eqnarray*}
p_{mfc}&=&\frac
{E(n_{mfc}^1)}{E(n_{mfc}^1+n_{mfc}^0)}=\frac{N_t^1 P(m, f, c | D=1)} {
N_t^1 P(m, f, c | D=1)+N_t^0 P(m, f, c | D=0)}
\\
&=&\frac{N_t^1 P(D=1|m,f,c)/P( D=1)} {
N_t^1 P(D=1|m,f,c)/P( D=1)+N_t^0 P(D=0|m,f,c)/P( D=0)}.
\end{eqnarray*}

This manipulation turns data from a retrospective design into a ``prospective''
likelihood stratified according to each type.
That is, the partial likelihood is the kernel of
$\operatorname{Binomial}(n_{mfc}$, $p_{mfc})$
and $\operatorname{Binomial}(n_{mc}$, $p_{mc})$,
and $n_{mfc}$ and $n_{mc}$ may be viewed as ``fixed'' as in a binomial
distribution. Thus,
although the second component of the likelihood
contains the parameters of interest,
it does not depend on the data.
It is also worth reemphasizing that, by modeling the counts using
partial likelihood instead
of the full likelihood, the dimensionality of the parameter space is
greatly reduced.
Only
phenocopy rate and genotype relative risks are estimated from the
partial likelihood.
Consequently, no assumption is needed regarding the underlying mating
type probabilities
in the population.

%s3 #&#
\section{Two existing methods for comparison}
%s3.1 #&#
\subsection{Constrained log-linear model}
A special case of our family case-control design, in which
$N_t^1=N_t^0=0$, was considered
in a recent study [\citet{casecontrol}] to detect maternal effects. They
used the log-linear
relative risk model for the case-mother and control-mother pairs, but
assuming the absence of
imprinting. Thus, their model is a special case of our model as
specified in (\ref{mm})
by setting $R_{{im}}=1$.
However, since their full likelihood approach requires the estimation
of mating type
probabilities, they propose to explore three levels of assumptions
regarding these
probabilities, namely, Mendelian inheritance, mating symmetry
and parental allelic exchangeability,
leading to what they call a constrained log-linear model (CLL).

In addition to the assumption about mating type probabilities,
the expected frequencies of control-mother pairs
were assumed to be the same
as the expected frequencies of any child-mother pairs for each type
regardless of whether the child is affected or not.
%This assumption is made apparent when one contrasts their expected
%frequencies of
%control-mother pairs (reproduced as the last column in our Table
%with the last
%column of our Table~\ref{tabl1} (bottom segment).
Considering the retrospective nature of the case-mother/control-mother
design, the fact that the child in a control-mother pair is unaffected
should be taken into account by multiplying the unaffected probability
by the expected frequency of each $(M,C)$ combination. Unless the
conditional probability of being unaffected given every $(M,C)$
combination is close to 1, approximating the frequencies of
control-mother pairs using the frequencies of child-mother pairs in the
general population would be inaccurate.
\citet{casecontrol}
justified the use of this approximation under the rare disease assumption.
This assumption only implies that the marginal unaffected probability
in the whole
population is close to 1, but the conditional unaffected probability given
some of the $(M,C)$ type may not be close to 1 at all. In particular,
when $(M,C)=(2,2)$, the conditional probability that the child is
unaffected, $1-\delta S_2R_2$, might be much smaller than 1 for
reasonably large relative
risks $S_2$ and $R_2$. For example,
suppose the phenocopy rate $\delta=0.05$, and the relative risks
are $S_2=2$ and $R_2=3$ as in one of
the settings in \citet{casecontrol}, then
$1-\delta S_2R_2$ is only 0.7.
Thus, although the rare disease assumption is necessary, it is
certainly not sufficient in
justifying the frequency approximation used in CLL.

%With respect to the three levels of assumptions on the mating type
%probabilities, Shi et~al. demonstrated that there is slight power gain
%even if one only imposes the constraint arising from the Mendelian
%inheritance, which universally holds for most human genes.

A careful dissection of CLL reveals that the
control-mother expected frequencies approximation was also needed for
the mating type probability assumption under Mendelian inheritance.
It is indeed true that, under Mendelian inheritance, the sum of the
expected frequencies of child-mother pairs with $(M,C)=(1,0)$ and with
$(M,C)=(1,2)$ equals the expected frequency with $(M,C)=(1,1)$ if the
pair is randomly sampled from the population. However,
this does not hold for control-mother pairs due to the involvement of
unaffected probabilities as discussed above, which complicates the
simple relationship in the constraint. %As for the mating symmetry and
%allelic exchangeability assumptions, since the mating selections are
%guided by cultural values and social rules in general, they are easily
%violated if there is any gender-specific assortative mating in the
%population.
%Simulation carried out in \citet{casecontrol} demonstrated slight power
%gain for
%testing
%association if these assumptions holds; however, biases on the
%parameter estimates
%and type I error rate if
%the assumptions are violated were not explored.

%s3.2 #&#
\subsection{Log-linear likelihood ratio test}

The log-linear likelihood ratio test (LL-LRT) was the first method
proposed for detecting imprinting and maternal effects simultaneously
using case-parent triad data
[\citet{Weinberg1998}]. %\citep{weinberg1998}.
It assumes multiplicative relative risks and models the counts of the
15 case-parent triads using a log-linear model as in (\ref{mm}). %(i.e.,
%Poisson regression).
It is necessary to make the assumption of mating symmetry when using
LL-LRT; otherwise, the number of parameters in the model would exceed
the degrees of freedom of the count data.

%s4 #&#
\section{Simulation}
To evaluate the performance of the proposed method and to compare it
with the existing methods, we create 8 simulation settings of the
relative risks due to variant allele, imprinting and maternal effects
(Table~\ref{tabl2}). The first 4 settings with the imprinting relative
risk being
equal to 1 (i.e., no imprinting effect) are the same as the simulation
settings in \citet{casecontrol}. Settings 5 and 6 have paternal and
maternal imprinting effects, respectively, but no maternal effect.
Settings 7 and 8 have paternal and maternal imprinting effects,
respectively, and also maternal effect. Prevalence of the disease is
set to be 0.05 (rare) or 0.15 (common). Note that the summation over
the 15 joint probabilities $P(D=1,M,F,C)$ equals the disease prevalence
$P(D=1)$, and thus the phenocopy rate can be solved from the equation
since relative risks and prevalence are set in our simulation and the
phenocopy rate is the only unknown quantity.

%
%t2 #&#
%
\begin{table}
\tablewidth=196pt
\caption{Simulation settings of the relative risks}\label{tabl2}
\begin{tabular*}{\tablewidth}{@{\extracolsep{\fill}}lcccccccc@{}}
\hline
&\multicolumn{8}{c@{}}{\textbf{Settings}}\\[-4pt]
& \multicolumn{8}{c@{}}{\hrulefill}\\
\textbf{Parameter} & \textbf{1} & \textbf{2} & \textbf{3}
& \textbf{4} & \textbf{5} & \textbf{6} & \textbf{7} & \textbf{8}\\
\hline
$R_1$&1&2&1&1&1&3&1&3\\
$R_2$&1&3&3&3&3&3&3&3\\
$R_{{im}}$&1&1&1&1&3&$1/3$&3&$1/3$\\
$S_1$&1&1&1&2&1&1&2&2\\
$S_2$&1&1&1&2&1&1&2&2\\
\hline
\end{tabular*}
\end{table}

We consider three variant allele frequencies (VAF), 0.1, 0.3 and 0.7,
and simulate the parental genotypes under two scenarios. Under the
first scenario, the population is
in HWE and the probabilities of a paternal or maternal genotype score
being 0, 1 and 2 are $(1-p)^2$, $2(1-p)p$ and $p^2$, respectively,
where $p$ is the variant allele frequency, $p\in\{0.1,0.3,0,7\}$. Since
the population is in HWE,\vadjust{\goodbreak} allelic exchangeability (AE) and mating
symmetry (MS) are implied. In the other scenario, the probabilities of
a genotype score being 0, 1 and 2 are $(1-p)^2(1-\zeta)+(1-p)\zeta$,
$2p(1-p)(1-\zeta)$ and $p^2(1-\zeta)+p\zeta$, respectively, where $\zeta
$ is the inbreeding parameter [\citet{Weir1996}], which is set to be 0.1 and
0.3 for males and females, respectively. As such, neither AE nor MS
holds in the population under the second scenario.

For each triad, the genotype of the child is sampled according to the
transmission probability given the previously simulated parental
genotypes, and then the disease status is generated as a Bernoulli
trial with the success probability being equal to the phenocopy rate
multiplied by the relevant relative risks. This process is repeated
until 150 case-parent triads and 150 control-parent triads are
obtained. This sample size was chosen to be the
same as in \citet{casecontrol} to facilitate comparison.
We then count the 15 types of child-parent triads among case families
and control families. Ignoring all the fathers, we also count the 7
types of child-mother pairs among case families and control families.
Finally, we also create a mixture sample of triads and pairs by
randomly setting the father's genotype to be missing with probability
0.5 in each triad. This last setting allowing for missing paternal
genotype is the most realistic in
genetic epidemiology studies. The number of replication for each simulation
setting is set to be 1000.\looseness=-1

LIME is applied to the pair/triad mixture counts (LIME-mix).
CLL is applied to the pairs (150 case-mother pairs and 150
control-mother pairs) with the constraint arising from AE.
Since CLL models pair counts and does not consider the imprinting
effect, LIME is also applied to pair counts without the imprinting
effect (LIME-pair) purely for the purpose of comparison with CLL.
On the other hand, LL-LRT is applied to the case-parent triads (150 complete
case-parent triads). The simulation results are summarized in the
following three subsections.

%s4.1 #&#
\subsection{Bias}
\begin{figure}

\includegraphics{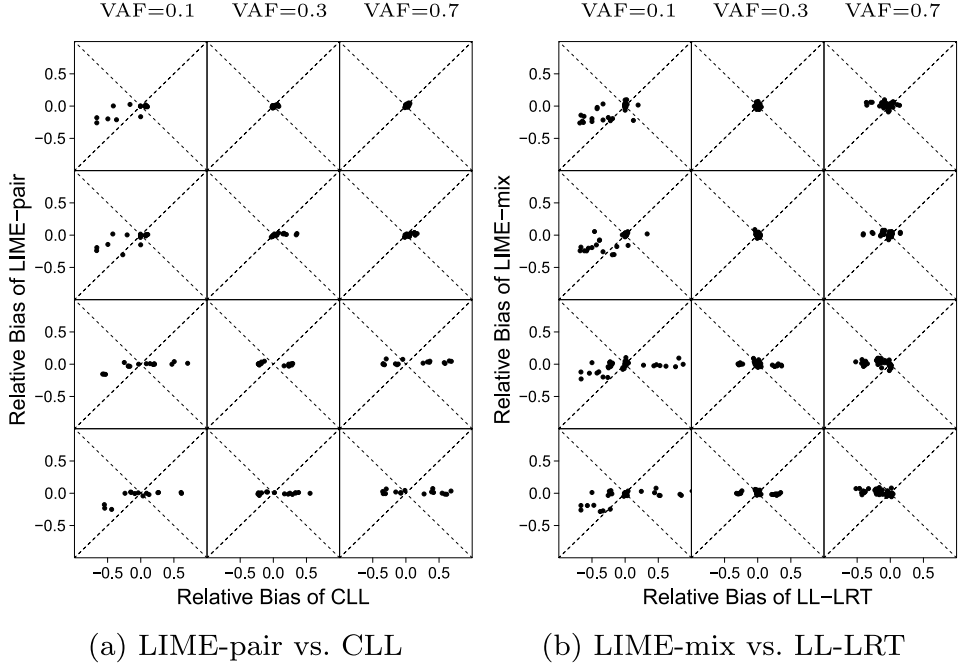}

\caption{Relative estimation biases of \textup{(a)}
LIME-pair versus CLL and \textup{(b)} LIME-mix versus LL-LRT. In both
figures \textup{(a)} and \textup{(b)}, three variant allele frequencies (VAF) are
considered and plotted in the three columns. Further, population 1 is
in HWE, such that both AE and MS hold, whereas neither AE nor MS holds
in population 2. The four rows in the figures correspond to the four
combinations of population (1 or 2) and disease prevalence (rare: 0.05
or common: 0.15): row~1: population${}={}$1, prevalence${}={}$0.05; row~2:
population${}={}$1, prevalence${}={}$0.15; row~3:
population${}={}$2, prevalence${}={}$0.05; row~4: population${}={}$2,
prevalence${}={}$0.15.}\label{figur1}
\end{figure}

%Issue 1 Approximation
Relative bias of a parameter estimate is defined as $(\hat{\theta
}-\theta)/\theta$, where $\hat{\theta}$ is the estimate of $\theta$ and
$\theta\in\{R_1, R_2, R_{{im}}, S_1$, $S_2\}$
as given in the multiplicative relative risk model. The scatterplots
shown in Figure~\ref{figur1}(a)
are relative biases of LIME-pair versus relative biases of CLL. Since
neither CLL nor LIME-pair models imprinting effect, it is only
appropriate to compare their biases under the settings in which the
imprinting effect is absent. Hence,\vspace*{1pt} each panel in
Figure~\ref{figur1}(a)
plots the biases of $\hat{R}_1$, $\hat{R}_2$, $\hat{S}_1$ and $\hat
{S}_2$ under the first 4 simulation settings of the relative risks. The
columns of panels in Figure~\ref{figur1}(a)
correspond to the three variant allele frequencies, while the rows
correspond to the four combinations of the two prevalence settings
(0.05 and 0.15) and whether AE holds. Recall that AE is the constraint
imposed in CLL.\looseness=-1

%
%f1 #&#
%

The scatterplots shown in Figure~\ref{figur1}(b) are relative biases of
LIME-mix versus relative biases of LL-LRT. Since both LIME-mix and
LL-LRT model variant allele, imprinting\vspace*{1pt} and maternal
effects simultaneously, each panel plots the biases of $\hat{R}_1$,
$\hat{R}_2$, $\hat{R}_{{im}}$, $\hat{S}_1$ and $\hat{S}_2$ under all 8
simulation settings. Thus, there are more points in Figure
\ref{figur1}(b) than in Figure~\ref{figur1}(a). The rows of Figure
\ref{figur1}(b) correspond to the four combinations of the prevalence
settings and whether AE holds. Note that AE implies MS, an assumption
made in LL-LRT.

The intersecting dotted diagonal lines in each panel divide the square
into 4 triangular regions. Scattering points in the left and right
triangles correspond to the relative biases on the $y$-axis
(LIME-pair/LIME-mix) being smaller than the relative biases on the
$x$-axis (CLL/LL-LRT) in absolute magnitude. In both
Figures~\ref{figur1}(a) and (b),
almost all the points are scattering in the left/right triangle in each
panel, which indicates the smaller relative biases of LIME-pair and
LIME-mix under all circumstances simulated. When the AE/MS does not
hold [3rd and 4th rows of both Figures~\ref{figur1}(a) and (b)], the relative
biases of CLL/LL-LRT become larger, a~phenomenon also noted in
\citet{Sinsheimer2003}. In contrast, the relative biases of
LIME-pair/Lime-mix remain at around the same level, which is close to zero.

When AE actually holds, the relative biases of CLL become a bit larger
when the prevalence changes from 0.05 to 0.15, which reflects the rare
disease assumption being necessary (though not sufficient) for CLL to
be valid. When both the rare disease (prevalence${}={}$0.05) and AE
assumptions hold but the variant allele frequency is small (VAF${}={}$0.1),
CLL still has some large relative biases, which mainly correspond to
the estimates of $R_2$ and $S_2$. When the VAF is small, it is likely
that there is no child-mother pairs with $(M,C)=(2,2)$ in the sample.
The zero cell count due to small VAF leads to large estimation
variability for these two parameters. This can also be observed from
the large relative biases of LL-LRT in estimating $R_2$ and $S_2$ when
VAF is small, which are due to the zero counts for $(M,F,C) = (2,1,2)$
and $(2,2,2)$.

%s4.2 #&#
\subsection{Type I error rate and power}

\begin{figure}

\includegraphics{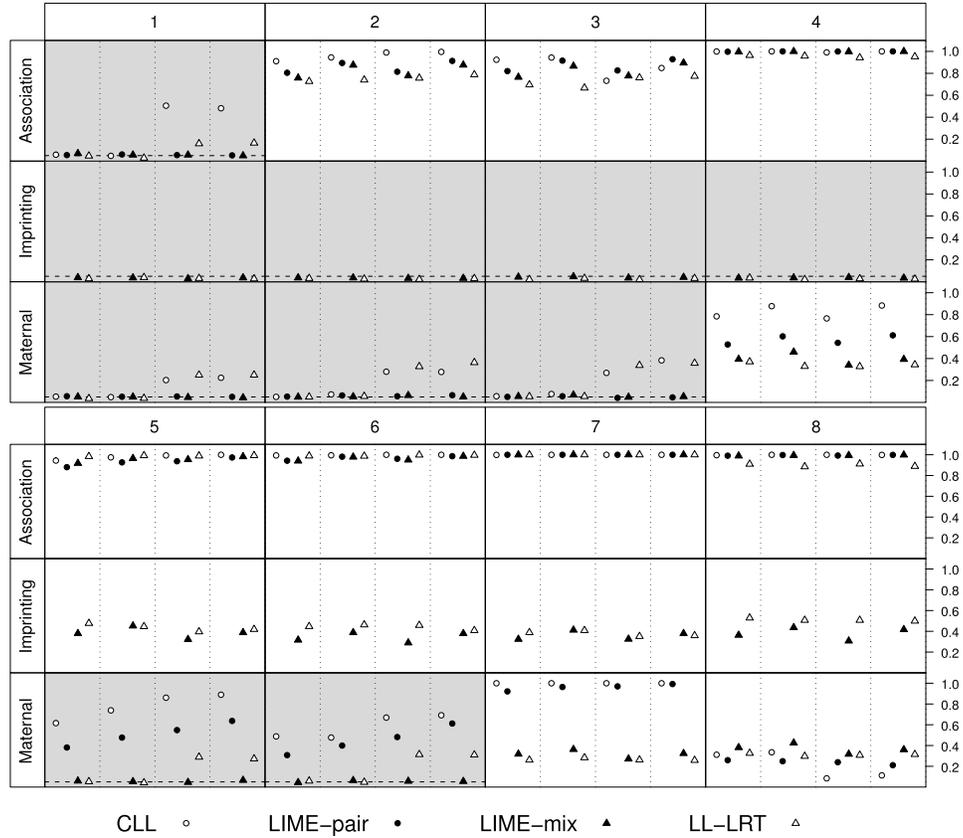}

\caption{Type I error rates and power of CLL, LIME-pair, LIME-mix and
LL-LRT. Panels with gray background give type I error rates, whereas
the rest show power. The vertical dotted lines divide each panel into 4
regions corresponding to, from left to right: (1) prevalence${}={}$0.05,
AE and MS holds; (2) prevalence${}={}$0.15, MS and AE hold; (3)
prevalence${}={}$0.05, neither AE nor MS holds; (4)~prevalence${}={}$0.15, neither AE nor
MS holds. Columns 1--8 correspond to the 8 simulation settings of
relative risks in Table \protect\ref{tabl2}.}\label{figur2}
\end{figure}

Figure~\ref{figur2} presents the type I error rates and power of CLL,
LIME-pair, LIME-mix and LL-LRT under the 8 simulation settings of
relative risks. The vertical dotted lines divide each panel into 4
regions, from left to right, corresponding to (1) prevalence${}={}$0.05 and
AE hold; (2)~prevalence${}={}$0.15 and AE hold; (3) prevalence${}={}$0.05 but
neither AE nor MS holds; (4) prevalence${}={}$0.15 but neither AE nor MS
%
%f2 #&#
%
holds. Type I error rates are shown in the gray panels with a
horizontal dashed line marking the nominal level of 0.05, while other
panels show power. Type I error rates and power of CLL and LIME-pair
are absent in the middle rows of the panels, since neither CLL nor
LIME-pair includes the imprinting effect.\looseness=-1

%s4.2.1 #&#
\subsubsection{Settings 1--4}
The imprinting effect is absent in simulation settings 1--4. Type I
error rates for detecting the imprinting effect using LIME-mix and
LL-LRT are all around the nominal level of 0.05. For detecting
association and maternal effect, CLL and LL-LRT have inflated type I
error rates in regions (3) and (4), in which neither AE nor MS holds,
whereas LIME-pair and LIME-mix have correct type I error rates under
all circumstances. The powers of CLL and LIME-pair are higher than
LIME-mix and LL-LRT, not surprisingly, since CLL and LIME-pair do not
attempt to estimate the nonexisting imprinting effect and
thus are more efficient. The power of CLL is higher than or about the
same as LIME-pair in regions (1) and (2), in which AE holds and is
correctly incorporated into CLL as a parameter constraint. However, the
power of CLL is lower than that of LIME-pair in regions (3) and (4)
under setting 3, a case in which the power for CLL is reduced because
the constraint is wrongly imposed when the assumption does not hold.
CLL has higher power than LIME-pair in setting 4 for detecting
maternal effect. The power of LIME-mix is always higher than that of
LL-LRT in detecting association and maternal effect in these four settings.

%s4.2.2 #&#
\subsubsection{Settings 5--8}
The imprinting effect is present in the simulation settings 5--8, and
maternal effect is present in the latter two settings. Under settings 5
and~6, CLL and LIME-pair have a lot of false positives for maternal
effects that are actually due to the imprinting effect because
of confounding [\citet{mimic}]. LL-LRT also has inflated type I error
rates in regions (3) and (4) due to its MS assumption being violated,
whereas LIME-mix has correct type I error rates in all regions. Due to
the strong association signal in these four settings, all four methods
have about the same power in detecting association. LL-LRT has higher
power in detecting the imprinting effect than LIME-mix, but LIME-mix
has higher power in detecting maternal effect than LL-LRT. Due to the
confounding between imprinting and maternal effect, the paternal
imprinting in setting 7 ``magnifies'' the signal of maternal effect if
imprinting effect is not properly accounted for. As such, CLL and
LIME-pair have a higher power for detecting the maternal effect, but
they miss the imprinting effect completely. On the other hand, setting
8 depicts maternal imprinting, and thus CLL and LIME-pair have a lower
power for detecting the ``reduced'' maternal effect. This represents
the worse case scenario; not only is the imprinting effect missed
completely but there is also reduced power for detecting maternal
effect.\looseness=-1

%s4.3 #&#
\subsection{Sensitivity analysis}
\label{ccsensitivity}
\begin{figure}

\includegraphics{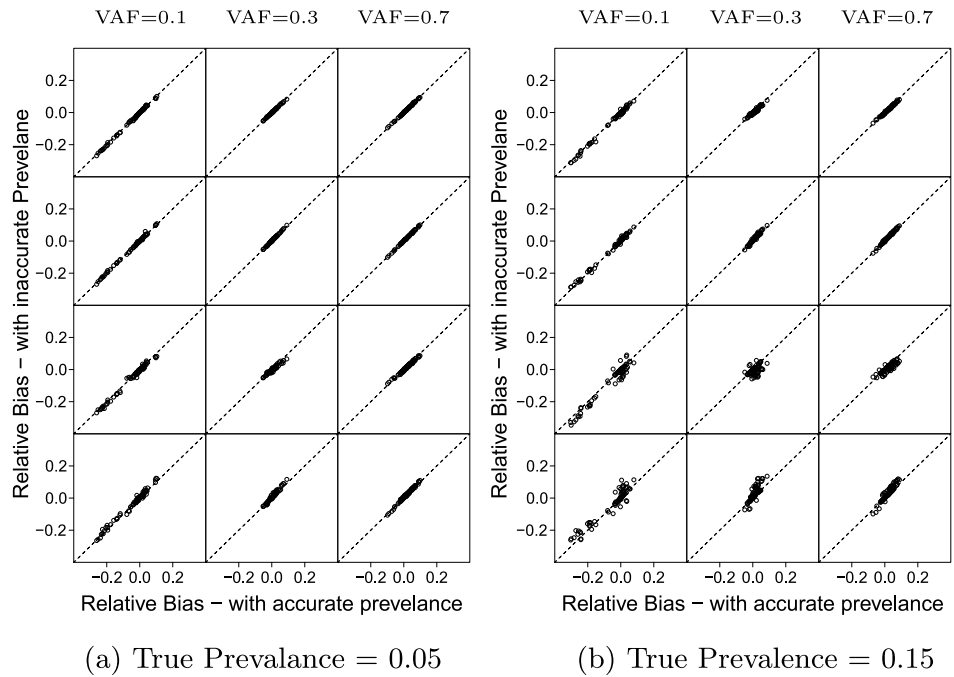}

\caption{Sensitivity analysis of the bias with a misspecified
prevalence when
\textup{(a)} the true prevalence${}={}$0.05 (rare disease) and \textup{(b)} the true
prevalence${}={}$0.15 (common disease). In both figures \textup{(a)} and \textup{(b)}, three
variant allele frequencies (VAF)
are considered and plotted in the three columns.
The four rows correspond to four different misspecified prevalences.
Row 1:
prevalence is 5\% higher than the true value; row 2:
prevalence is 5\% lower than the true value; row 3:
prevalence is 20\% higher than the true value; row 4:
prevalence is 20\% lower than the true value.}\label{figur3}
\end{figure}

Since the proposed method LIME-mix needs the disease prevalence in the
population as a known constant, we conduct a sensitivity analysis to
investigate the impact of using a misspecified prevalence in our model.
A constant that is 5\% higher, 5\% lower, 20\% higher or 20\% lower
than the true prevalence in the simulation setting is used as the
prevalence in LIME-mix. The mixtures of triads/pairs simulated under
the 8 settings of relative risks are analyzed again using LIME-mix with
these inaccurate prevalences. Relative biases of LIME-mix using
inaccurate prevalence are plotted versus relative biases of LIME-mix
using the true prevalence in Figure~\ref{figur3}.

%
%f3 #&#
%

When the disease is rare (with a true prevalence of 0.05), we can see
from Figure~\ref{figur3}(a)
that all points fall almost exactly on the dashed line with slope 1,
indicating that the estimates of the relative risks using the
inaccurately specified prevalence are actually
very close to those using the true prevalence. That is, the results are
rather insensitive
to the misspecification of population prevalence. On the other hand,
when the disease
is more common (true prevalence being 0.15), the points still scatter
around the
dashed line with slope 1, as we can see from Figure~\ref{figur3}(b), although
there is more
scattering when the
prevalence is incorrectly specified to be 20\% higher/lower than the
true value.
Overall, the results are reasonably insensitive to the specification of
population
prevalence. Estimated prevalences are often obtained through large
scale population studies and would, in those cases, not deviate greatly
from their true values. As such, our results seem to suggest that LIME
would give reasonably accurate estimates by using population
prevalences estimated from external sources when the same population
and diagnostic criteria have been studied.

%s5 #&#
\section{Discussion}
%Novelty
In this paper we propose a partial likelihood approach for detecting
imprinting and
maternal effects simultaneously using case-control families.
The crucial role played by imprinting and maternal effects in complex
human diseases has
been increasingly explored in the last few years, as simply focusing on
sequence variation
has been proven to be insufficient for studying disease etiology.
As such, there is an immediate need for robust and efficient
statistical methods
for detecting imprinting and maternal effects, and our contribution is
one step in this direction.
Our proposed method possesses a number of novel features compared to
existing ones in
the literature. We augment the traditional case-parent triads design to
a family-based
case-control design by
recruiting control-parent triads as well.
This differs from \citet{Weinberg2005} in that they only genotype the
parents of the controls, leading to a case-parent triad/control-parent design.
Further, we allow for
missing fathers considering the fact that fathers are often harder
to recruit in family-based studies. Thus, this design also differs from
that of \citet{Vermeulen2009},
as the current design also recruits control-parent triads and
case-mother pairs.
By recruiting control families of the same structure as case families,
we create ``internal matches'' stratified by the familial genotypes. As such,
we can extract from the full likelihood of the retrospective design a partial
likelihood component that can be thought of as the products of likelihoods
from stratified prospective designs.
Through this conditional on the familial genotypes, the nuisance parameters
with respect to the population mating type probabilities (i.e.,
probabilities of parental
genotype combinations) are no longer involved in the partial likelihood.
As such, it is no longer necessary to make any assumption
about mating type probabilities; such assumptions are strong and
usually unrealistic but are
made in the literature in an effort to reduce the number of parameters
for the
full likelihood approach.
This alleviates the problem of over-parametrization that often plagues
existing methods.
Furthermore, our formulation takes into account the fact that control
families have unaffected children, making it applicable to both rare
and common diseases as opposed to
just rare diseases in some existing methods.
However, we note that the mother-child data type in which both the
mother and the
child are heterozygotes cannot be included in the analysis, an issue
that will be
discussed further below.

%Results
Through simulation with a variety of settings, including some adopted
from the literature,
we demonstrate the robustness of LIME to violation of assumptions
on absence of imprinting effects, rarity of disease and mating type
probabilities.
First, by utilizing a mixture of case/control triad families and case/control
pair families, imprinting and maternal effects can be studied jointly
to address the confounding issue faced by approaches that
assume the absence of imprinting effect when detecting maternal effects.
As such, LIME
has correct type I error rate compared to those approaches.
If, however, there is a priori and unequivocal information that
imprinting effect is indeed absent,
then a method that assumes the absence of imprinting effect will
usually lead to gain in power for detecting
maternal effect.
In this situation, a version of LIME that assumes the absence of
imprinting (by setting $R_{{im}}=1$) is recommendable
given its robust and efficient features.

Second, regardless of whether the disease is common or rare, LIME has
very little bias
in the estimates of the model parameters. In contrast, CLL, which
assumes the disease
being rare, has much larger biases, even when the disease is indeed rare,
since rare disease is a necessary but not a sufficient condition for
CLL to be valid.

Finally, mating symmetry is commonly assumed for many imprinting and/or maternal
effects detection methods [\citet{Ainsworth2011},
\citet{casecontrol}, \citet{Weinberg1998},
\citet{Weinberg1999}, \citet{jiyuanCI}]. However, when this assumption is violated, there
can be large biases and greatly inflated type I error rates,
whereas LIME is not affected at all by departure from such assumptions
(Figures~\ref{figur1} and~\ref{figur2}).
In particular, when there is population substructure, the assumption of
HWE is violated
even if there is HWE in each of the subpopulations. Therefore,
population substructure will likely
exert a large effect on methods that assume some mating frequency
distributions. In contrast, because
the partial likelihood is independent of the mating type parameters,
LIME may not be as sensitive to
such population substructure.
% but what if the disease prevalences also differ in the subpopulations?
The non-HWE scenario considered in our simulation may be viewed as a
mixture of
two subpopulations, one is in HWE and the other is inbred.
Indeed, CLL and LL-LRT are highly sensitive, whereas LIME is robust to
this type of
population stratification. However, will LIME still be robust if
the disease prevalences also differ in the subpopulations?
To investigate this, we consider the following three scenarios of
population substructure in which a population consists of two distinct
subpopulations each in HWE: (a) the two populations differ in their
allele frequencies (VAF${}={}$0.1 or~0.3)
but with the same disease prevalence of 0.05; (b) the two populations
have the same allele frequency (VAF${}={}$0.3) but different
prevalences (0.05 or 0.15); (c) the two populations have different
allele frequency
(VAF${}={}$0.1 or 0.3) and
also different prevalences (0.05 or 0.15).
Under disease risk settings 1--4 (Table~\ref{tabl2}; imprinting
absence), LIME has
smaller relative biases in all parameter estimates and smaller type I
error rates
under all scenarios, and they are smaller (some are much smaller) than
those from CLL.
For disease settings 5--8 (Table~\ref{tabl2}; imprinting present),
LIMEs estimates have larger biases, especially for scenario (b), but
they are, in the
majority of
cases, much smaller than those from CLL. This indicates that LIME is sensitive
to population stratification when there is imprinting effect, but it is
certainly
more robust than the alternative method even under the situation where the
disease prevalences are different in addition to different frequencies
in the
subpopulations.

%Power
%Need to work on this again after reviewing new Table~\ref{tabl3}

%The major advantage of CLL as claimed by the authors \cite{casecontrol}
%is its ability
%to exploit the information contained in the mating type probabilities
%to achieve greater
%power. Although CLL generally results in greater power than LIME-pair,
%the difference
%is usually small and in the expense of inflated type I error and bias
%when the assumption
%is violated. Further, there are cases in which LIME will not only have
%correctg type I
%error but also greater power as well (Figure~\ref{figur3} fields 3 and
%4 of
%setting 3). This is
%due to the elimination of the need to estimate the nuisance parameters
%and thus smaller
%variability of the estimates of the paramters of interest.

%Limitation

To compare with the performance of CLL, we consider LIME-pair in
addition to LIME-mix.
In real data applications, only LIME-mix will be used unless there are
no triads in the
sample (i.e., all the fathers are missing in all the families). This
is an unlikely
scenario, but if this does happen, then LIME-pair is recommendable over
CLL due to its
robust feature, although we note that LIME-pair, like CLL, also assumes
the absence of imprinting
effect.\looseness=-1

%Limitation 2
The need to specify disease prevalence deserves further discussion.
Since the prevalence is in fact a function of the disease model
parameters and mating
type probabilities, it would be attractive if one can make use of this
fact without the need
to specify this parameter. However, in order to use the current
argument, the
partial likelihood needs to be free of the mating type probabilities,
which will no longer
be true if we express $P(D)$ in terms of the other parameters.
Fortunately, as we have pointed our earlier,
estimated prevalences are often obtained through large scale population
studies and would
usually not deviate greatly from their true values.
Overall, even if the prevalence is
misspecified as much
as 20\% away from the true value, the estimates are still quite close
to those had the
prevalence been specified correctly, reaffirming the robustness of the
proposed procedure.
However, for rarer diseases, the deviations between the true and
estimated prevalences
may be larger.
Thus, we further investigate the effect of greater prevalence
misspecification for a disease with a prevalence of 1\%.
We use a prevalence of 0.2\%, 5\% (representing 500\% decrease or
increase) and 0.8\%, 1.2\% (20\% decrease or increase).
Our results show that LIME remains robust with a 20\% increase/decrease
misspecification. Even with a 500\% decrease, the parameter estimates
closely track
those using the true prevalence. A 500\% increase leads to more sensitivity,
but the procedure is still quite robust if the variant allele frequency is
moderate.

In LIME, the mother-child data type in which both the mother and the
child are heterozygous---the $(1,1)$ data type---is not included in the analysis
since the nuisance parameters are tangled up with the risk parameters
of interest (Table~\ref{tabl1}), and therefore its inclusion will
render it impossible to adopt the partial likelihood approach. This
exclusion, however, will lead to loss of information.
Since we cannot evaluate its effect on LIME directly, we use CLL as a
surrogate by analyzing the data without the $(1,1)$ data
type and compare the results with those from the earlier analysis where
$(1,1)$ is included. The power is slightly lower for most
of the tests, although power actually increases for some; overall, the
average power drop is 0.056, small but appreciable.
This amount may be viewed as an upper bound; the effect on LIME is
expected to be much smaller unless most of the data are from pairs, not
triads. This finding of small effect is not surprising
considering the parent asymmetry test (PAT) of \citet{Weinberg1999}.
Therein, five triad categories, including $(1,1,1)$ (all three people in
the family are heterozygous),
were omitted from the model, but PAT is competitive in power for
imprinting test compared to methods that use all categories.

%Limitation 3
Although fathers are usually harder to recruit than mothers, this does
not imply
100\% participation rate from mothers. As such, there will likely be
child-father
pairs in a study as well, but such data cannot be utilized in our
partial likelihood
approach since the model parameters and nuisance parameters cannot be factored
as in child-mother pairs.

%robus
%robustness to sampling zero

%limitations
%Families with multiple affecteds
Finally, in this paper we assume a single affected child for each
case family. In genetic diseases, there may be familial aggregation and,
therefore, it is not unlikely that there may be multiple affected siblings
in a family. If each family is being recruited through single
ascertainment (i.e., through an affected/unaffected child), then additional
information from the siblings (even from the control families)
may be used in the partial likelihood formulation.
This may lead to substantial power gain, and thus warrants further
investigation.

\section*{Acknowledgments}

We would like to thank the Editors and two anonymous referees for their
constructive comments and suggestions, which have led to improved
presentation and content of the paper.

%suskaldyti doi

% imsref loaded by lrinkeviciute, 2012-08-31 11:00:31
% imsref loaded by lrinkeviciute, 2012-08-31 11:13:21
% imsref loaded by lrinkeviciute, 2012-08-31 11:16:25

\printaddresses

\end{document}